\documentclass[
    ,final            
  ]
  {aipproc}

\layoutstyle{8x11double}

\begin{document}

\title{Line Widths of Single--Electron Tunneling
Oscillations: Experiment and Numerical Simulations}



%
%
%
%


\author{Jonas Bylander, Tim Duty and Per Delsing}{
  address={Dept.\ of Microtechnology and Nanoscience,
Chalmers University of Technology, SE--412 96 G\"{o}teborg,
Sweden} }

%

\begin{abstract}
We present experimental and numerical results from a real-time
detection of time-correlated single-electron tunneling
oscillations in a one-dimensional series array of small tunnel
junctions. The electrons tunnel with a frequency $f=I/e$, where
$I$ is the current and $e$ is the electron charge. Experimentally,
we have connected a single-electron transistor to the last array
island, and in this way measured currents from 5~fA to 1~pA by
counting the single electrons. We find that the line width of the
oscillation is proportional to the frequency $f$. The experimental
data agrees well with numerical simulations.

%
%
%
\end{abstract}

\maketitle


\section{Introduction}

We have recently reported a measurement of a very small electrical
current, $I$, (5 fA--1 pA) by direct counting of the single
electrons that tunnel through a one-dimensional series array of
metallic islands separated by small tunnel junctions
\cite{Bylander-Nature-05}. At low temperature, and when the array
junction resistance is greater than the Klitzing resistance,
$R_\mathrm{K}\approx 26~\mathrm{k}\Omega$, charge is localized and
an excess electron charge, $e$, on one island polarizes the
neighboring islands, thus repelling other electrons
\cite{Likharev-IEEETransMag25-89,Bakhvalov-SovPhysJETP-89}. This
potential profile, extending over a distance of
$M=\sqrt{C_\mathrm{A}/C_0}$ islands, is often called a "charge
soliton"; here $C_\mathrm{A}$ and $C_0$ are the junction
capacitance and island stray capacitance, respectively.
Electrostatic repulsion creates a lattice of solitons in the
array, which above a certain threshold voltage starts to move as a
whole, leading to an oscillation of the potential of any island
inside the array with the mean frequency
\begin{equation}\label{fIe}
f=I/e.
\end{equation}

In \cite{Bylander-Nature-05}, the current was injected through a
tunnel junction from the array into the island of a
single-electron transistor (SET)
\cite{Likharev-IEEE-mag-86,FultonDolan-PRL-87}, a sensitive
electrometer. The SET was embedded in a resonant $LC$ circuit and
operated in the radio-frequency mode (RF-SET)
\cite{Schoelkopf-Science-98}. By monitoring the output signal from
the RF-SET, we were able to detect the single-electron tunneling
events in real time, and the power spectrum showed a peak at the
frequency (\ref{fIe}), thus demonstrating time correlation.

This type of electron counter can potentially be used to measure
small currents with very good accuracy, and without any need for
calibration since the only parameter involved in (\ref{fIe}) is a
natural constant.

In this paper we compare the measured spectral line widths with
those obtained from a computer simulation.

\section{Experimental techniques}
We made the device out of Al/AlOx with an SET resistance of
$R^\mathrm{SET}\!=\!30~\mathrm{k}\Omega$ and charging energy
$E_C^\mathrm{SET}/k_\mathrm{B}\!=\!1.6$~K, where $k_\mathrm{B}$ is
Boltzmann's constant. We obtained a small-signal charge
sensitivity of $\delta q\!=\!2\!\cdot\!
10^{-5}~e/\mathrm{Hz}^{1/2}$ and a bandwidth of $10$~MHz at an RF
carrier frequency of $358$~MHz. The $N=50$ junction array, with
junction parameters
$R_\mathrm{N}^\mathrm{Array}\!=\!940~\mathrm{k}\Omega$,
$C_\mathrm{A}\!=\!0.42$~fF and $C_0\!=\!0.030$~fF, had the
charging energy $E_C^\mathrm{Array}/k_\mathrm{B}=2.2$~K per
junction, and a soliton size of $M\approx 3.7$ islands (that is,
$N\gg M$).

We performed the experiments in a dilution refrigerator at
$0.03$~K and a parallel magnetic field of $475$~mT. The device was
superconducting, but with a suppressed gap,
$\Delta/k_\mathrm{B}\approx 0.6$~K. At this field, the Josephson
coupling energy is very small in our junctions:
$E_\mathrm{J}/k_\mathrm{B}<10~\mathrm{mK}\ll T$. Therefore Cooper
pair tunneling is strongly suppressed, and we see only
quasiparticles. Furthermore, we take advantage of the subgap
resistance, $R_\mathrm{SG}=\mathrm{e}^{\Delta/k_\mathrm{B}T_e}
R_\mathrm{N}^\mathrm{Array}\approx 50
R_\mathrm{N}^\mathrm{Array}$, for an electron temperature
$T_e\approx 0.15$~K, estimated using parameters for
electron--phonon coupling and Kapitza resistance
\cite{Verbrugh-JAP-95}. This makes the onset of current more
gradual than in the normal state, thus reducing the sensitivity to
bias voltage fluctuations.


\begin{figure}
  \includegraphics[height=.25\textheight]{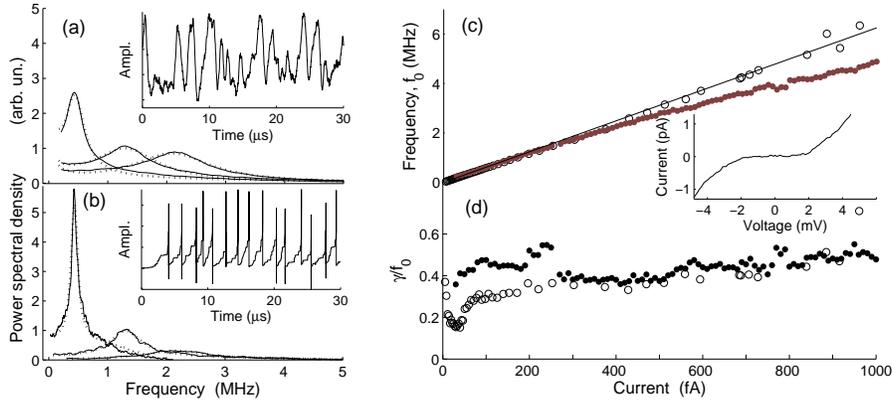}
\caption{Experimental (a) and simulated (b) power spectral
densities   (solid lines) with Lorentzian fits (dotted). The
currents   were 79, 220 and 370~fA (a), and 71, 221, 374~fA (b).
Insets: Time series for $I=80$~fA (in (b) before filtering), where
each peak represents
  one electron passing through the SET. Comparison of fitted
  frequencies, $f_0$, (c) and relative line
  widths, $\gamma/f_0$, (d) for experimental (dots) and simulated
  (circles) spectra. Solid line in (c): $I=ef$.
  Inset in (c--d): Array $I\!-\!V$ curve at $B_{||}=475$~mT.}
  \label{figure}
\end{figure}

\section{Numerical methods}
We have numerically simulated the electron transport in the array
using a direct Monte Carlo method
\cite{Likharev-IEEETransMag25-89,Bakhvalov-SovPhysJETP-89}. We
calculate the tunnel rates in each junction by using the
"orthodox" theory of single electron tunneling
\cite{AverinLikharev-MesoPhenomSolids}, where the probability of a
tunneling event per unit time is fully determined by the change in
free energy of the system. The time step between such events is
determined by a random variable that mimics the stochastic nature
of tunneling. Below the superconducting gap, $V<2\Delta/e$, we
have phenomenologically introduced the subgap resistance,
$R_\mathrm{SG}$. For simplicity, however, we disregard the
superconducting density of states above the gap.
The simulations are idealized in the way that we disregard
disorder and external noise, as well as bias fluctuations or
drift.

We model the charge detection by sampling the potential of the
last array island at a rate corresponding to the bandwidth of the
$LC$ circuit.

\section{Results and discussion}
From the experimental data we have subtracted a background to
account for amplifier and external noise, including an approximate
$1/f$ term. Then, both experimental and numerical power spectra
are fitted to the function,
\begin{equation}\label{lorentzian}
S(f) = a\gamma^2/\left( (f-f_0)^2+\gamma^2 \right),
\end{equation}
a  Lorentzian around a center frequency $f_0$, with a half width
$\gamma$, see Fig.~\ref{figure}(a--b). We find that $f_0$ agrees
with (\ref{fIe}), see Fig.~\ref{figure}(c), and that $\gamma$ is
proportional to the current: $\gamma/f_0\approx 0.4$, see
Fig.~\ref{figure}(d).

For $I<200$~fA, the measured peak is broad compared to the
simulation, likely due to the difficulty in maintaining a stable
bias for very small currents. We estimate the noise from the
biasing circuitry and thermal EMF to be of the order of $10~\mu$V,
causing a current fluctuation of a few fA during the measurement
time.
The deviation from (\ref{fIe}) in Fig.~\ref{figure}(c) for
$I>400$~fA arises probably because of an increasing asymmetry of
the peak, which causes the fit (\ref{lorentzian}) to underestimate
the frequency.

In conclusion, we have shown that the line width of the
single-electron tunneling oscillations is proportional to the
frequency, and that the line width extracted from our simulations
agrees well with the experimental data.


\begin{theacknowledgments}
We thank K. Bladh, D. Gunnarsson, S. Kafanov, M. Taslakov and C.
Wilson for assistance and discussions.
\end{theacknowledgments}



\bibliographystyle{aipproc}   

\bibliography{references}


\end{document}